\documentclass{article}
\usepackage{latexsym}
\usepackage{graphicx}
\usepackage{epsfig}
\newcommand \be {\begin{equation}}
\newcommand \bea {\begin{eqnarray}}
\newcommand \ee {\end{equation}}
\newcommand \eea {\end{eqnarray}}
\newcommand{\Z}{\mathbf{Z}}

\begin{document}
\topskip 2cm
\begin{titlepage}
\rightline{\today}

\begin{center}
{\Large \bf 
Renormalization group analysis of the three-dimensional Gross-Neveu model at finite temperature and density} \\
\vspace{2.5cm}
{\large Paolo Castorina$^{1,2}$, Marco Mazza$^{1}$, Dario Zappal\`a$^{2,1}$} \\
\vspace{.5cm}
{{\sl $^{1}$ Department of Physics, University of Catania} \\
{\sl Via S. Sofia 64, I-95123, Catania, Italy}\\
\vspace{.2cm}
{\sl $^{2}$ INFN, Sezione di Catania,  Via S. Sofia 64, I-95123, Catania, Italy}} \\
\vspace{1,cm}
{\sl e-mail: paolo.castorina@ct.infn.it;  ;  dario.zappala@ct.infn.it}\\
\vspace{3.2cm}

\begin{abstract}
The Renormalization Group flow equations obtained by means of a proper time regulator are used to analyze the restoration of the discrete chiral symmetry at non-zero density 
and temperature in the Gross-Neveu model in $d=2+1$ dimensions. The effects of the wave function renormalization of the auxiliary scalar field on the transition  have 
been studied.  The analysis is performed for a  number of fermion flavors $N_f=12$ and the limit of large $N_f$ is also considered. The results are compared with those coming 
from  lattice simulations. 
\end{abstract}
\end{center}

\vspace{2.5cm}


\end{titlepage}

\newpage
\par

In recent years there has been a renewed interest in the chiral symmetry restoration in QCD in connection with its 
possible experimental observation in the transition from the confined to the  quark-gluon plasma phase.
Lattice simulations, which are the most relevant tool to study such nonperturbative phenomena, when employed to investigate  QCD at finite 
baryon density, must deal with the main problem of the appearance of complex terms in 
the Euclidean action due to the presence of the chemical potential and only recently new lattice techniques have been proposed \cite{fod,for},
which overcome this problem, at least for not too large values of the chemical potential.
On the other hand, the chiral symmetry breaking at finite temperature and density has been investigated in other models \cite{klim,hands1,hands2,hands3,hands4}, 
which have a well defined action even for nonvanishing chemical potential and which, despite their simplicity, give a qualitative picture of the transition that could 
provide reasonable indications of this phenomenon in the full QCD framework.

The simplest  model is represented by the Gross-Neveu model \cite{gross}, characterized by either  discrete or continuous chiral symmetry, which in three 
spacetime dimensions shows a critical coupling $g^*$ that indicates the threshold for the occurrence of chiral symmetry breaking at zero temperature and density.
This model has an interacting continuous limit with a spectrum that contains composite fermion and antifermion states analogous to baryons and mesons 
and its  four fermion interaction can be regarded as an effective interaction for quarks at intermediate energies
and at not very high temperatures where the effect of heavy mesons is still suppressed. Moreover in three dimensions the model is renormalizable 
order by order in the $1/N$ expansion around the critical coupling $g^*$ \cite{wil1}.

In addition to lattice calculations, another nonperturbative tool, the Exact Renormalization Group (ERG) 
(see \cite{wettrep} for a review of the subject and a more  complete list of references)
has been employed to study four fermion models, such as the Nambu-Jona-Lasinio model at finite temperature and density analyzed in \cite{wett1}, as well as the 
fixed point structure of the Gross-Neveu model in  \cite{wett2}.
In this letter we make use of the  Renormalization Group to investigate the specific problem of the phase diagram of the tridimensional Gross-Neveu model 
in order to explicitly test this approach through a comparison with the lattice analysis. We shall focus on  a particular version of the Renormalization Group 
flow equation, based on the heat kernel representation of the effective potential, which has already been used in similar contexts, namely for analyzing the 
four-dimensional linear sigma model at finite temperature and density\cite{pirner1,pirner2,deandrea}. 

This latter version of the flow equation, often indicated as Proper Time Renormalization Group (PTRG) is typically obtained as an improvement of the one loop 
effective action by making use of the proper time representation \cite{schwinger} and introducing a suitable  multiplicative cutoff function  as a regulator of the 
infrared modes \cite{ole,senben1}. The PTRG  does not belong to the class of ERG flows which can be formally derived  from the Green's functions generator 
without resorting to any truncation or approximation \cite{litim1,litim2,litim3}. Nevertheless the PTRG flow equation has the property of preserving the symmetries 
of the theory and, in addition, it has  a quite simple and manageable structure. Moreover the PTRG flow, although not able to fully reproduce perturbative 
expansions beyond the  one loop order \cite{zappa1},  provides excellent results when used to evaluate the critical properties such as the critical exponents
of the three dimensional scalar theories at the nongaussian fixed point \cite{boza,maza}. In particular the determination is optimized when one 
takes the sharp limit of the cutoff function on the proper time and it turns out to be much more accurate than the one corresponding to a smoother regulator.
The PTRG  also provides excellent determinations of the  energy levels of the quantomechanical double well \cite{zappa2}. These results are  obtained at 
the first order in the derivative expansion of the effective action, i.e. by reducing the full flow equation to two coupled differential equations for the potential 
and the wave function renormalization, and  in our opinion this is a clear indication that, within this specific approximation,  the PTRG flow is reliably accurate. 
According to this point of view it is interesting to test the PTRG flow in a different context which involves fermionic systems at finite temperature and density and 
in particular to compare the results coming from this approach  with those obtained from  the lattice investigations.

Therefore in the following we focus on the Gross-Neveu model with Euclidean action 
\be\label{action}
S^{(0)}=\int d^3 x \left ( i \overline \psi_j   \partial \hskip -0.2 cm \slash \psi_j - \frac{\widehat g^2}{2N_f}(\overline \psi_j \psi_j)^2 \right ) 
\ee
where $ \partial \hskip -0.2 cm \slash= \gamma^\mu \partial_\mu $ involves three gamma matrices ($\mu$ takes the three values:  $1,2,4$), $\psi_j$ is a four 
component spinor and  $j$ is the flavor index :  $j=1,\ldots N_f$.  
The action (\ref{action}) is invariant under the discrete chiral transformations $\psi_j \to \gamma^5 \psi_j$ ,  $\overline \psi_j \to - \overline \psi_j \gamma^5$.
 
As it is well known \cite{gross}, instead of considering Eq. (\ref{action}), it is helpful to introduce an auxiliary scalar field $\sigma$  and put the action in a bosonized form. 
In fact after adding to $S^{(0)}$ the  term  $\int d^3 x (1/2)(\sigma +\widehat g \overline \psi_j \psi_j /\sqrt{N_f})^2$, which is quadratic in $\sigma$ and therefore does not 
modify the fermionic theory, one alternatively describes the model by means of the action 
\be\label{actionone}
S=\int d^3 x \left ( \frac{1}{2} \sigma^2+ i  \overline \psi_j  \partial \hskip -0.2 cm \slash \psi_j +  \frac{\widehat g}{\sqrt{N_f}}\sigma \overline \psi_j \psi_j \right ) 
\ee
The action  (\ref{actionone}) is to be taken as the bare action which fixes the ultraviolet boundary conditions of the renormalization group flow. In fact in the flow equations 
we consider a more general  structure of the effective action which includes a scalar  potential $U_k(\sigma^2)$, the kinetic term for the scalar sector, properly normalized by 
the wave function renormalization $Z_k$ and also a fermionic  wave function renormalization $Z^\psi_k$: 
\be\label{effact}
S_k=
\int d^3 x \left ( \frac{Z_k}{2} \partial_\mu  \sigma \partial_\mu  \sigma +U_k(\sigma^2)+i Z^\psi_k  \overline \psi_j  \partial\hskip -0.2 cm \slash \psi_j
+\frac{\widehat g_k}{\sqrt{N_f}}\sigma \overline \psi_j \psi_j \right ) 
\ee
In Eq. ({\ref{effact}) we have explicitly introduced a dependence of the various terms on the  running momentum scale  $k$ which parametrizes the PTRG flow.  
In fact the renormalization group equations describe their evolution as functions of $k$,  starting from the 
ultraviolet boundary conditions given in Eq. (\ref{actionone}), down  to  the far infrared region,  when  all modes have been integrated out.  
So the ultraviolet boundary conditions, corresponding the large value $k=\Lambda$, are  $Z_\Lambda=0$, $Z^\psi_\Lambda=1$,  $U_\Lambda(\sigma^2)=\sigma^2/2$, 
and a fixed  value for $\widehat g_\Lambda$. We shall limit ourselves to field independent wave function renormalizations  $Z_k$ and $Z^\psi_k$ and to a potential  
which is polynomial in $\sigma^2$.

The shape of effective potential $U_k(\sigma^2)$  in the infrared region, i.e. at $k\sim 0$, does provide informations about the chiral symmetry breaking  and restoration, 
which are  in fact related to the expectation value of the auxiliary scalar field  $\sigma$. If this expectation value is nonvanishing then the fermion becomes massive 
through the Yukawa coupling in  (\ref{effact}) and the chiral symmetry of the bare action is spontaneously broken. 
As already mentioned, the mean field analysis indicates that at zero temperature and density the symmetry is broken only for couplings larger than a critical value, 
$\widehat g> \widehat g^*$, whereas below $\widehat g^*$ the field $\sigma$ has zero expectation value and the fermion is massless, preserving chirality. 
When  $\widehat g>\widehat g^*$ a sufficient  increase  of the temperature and (or) of the density is expected to restore the symmetry.
 
A saddle point expansion of Eq. (\ref{actionone}) yields the one loop effective action 
\be\label{olact}
S_{1l}=S-\frac{1}{2} {\rm Tr}~ {\rm ln} \left ( {\cal A}^\dagger {\cal A}\right)+
\frac{1}{2} {\rm Tr}~ {\rm ln} \left \lbrack \frac{\delta^2 S}{\delta \sigma\delta \sigma} -2  \frac{\delta^2 S}{\delta \sigma\delta \psi} 
{\cal A}^{-1} \frac{\delta^2 S}{\delta \overline \psi \delta \sigma}
\right \rbrack
\ee
where 
\be\label{matfer}
{\cal A}=\frac{\delta^2 S}{\delta \overline \psi \delta \psi}
\ee
The logarithms in Eq. (\ref{olact})  can be  expressed by means of the proper time representation 
\be\label{loga}
{\rm ln} (X) = -\int^\infty_0 \frac{d~s}{s} ~e^{-sX} 
\ee
and the  infrared modes are properly regularized by means of  a cutoff function $f_k$, \cite{senben1}, multiplicatively introduced into Eq. (\ref{loga})
by replacing $e^{-sX}$ with  $e^{-sX}~f_k$. 
Clearly this replacement  introduces a dependence on the variable $k$ in Eq. (\ref{olact}). Then the PTRG equations are obtained by replacing everywhere in 
Eq. (\ref{olact}) the action with the $k$ dependent ansatz $S_k$ of Eq. (\ref{effact}) and by taking the derivatives with respect to $k$ of each side of Eq. (\ref{olact}).
Typically $f_k$ is taken as a smooth heat kernel cutoff  \cite{senben1,pirner1,pirner2,deandrea}, however as shown in \cite{boza,maza} the PTRG 
equations are optimized when the sharp limit of  $f_k$ is taken, or, more precisely  when  its  derivative with respect to $k$ ( which is the relevant quantity in the 
differential flow equations) is taken as  a delta-function\cite{litim2,zappa1}
\be
\label{delta}
 k \partial_k f_k= -2s ~\delta \left ( s- \frac{1}{k^2} \right ) 
\ee
Therefore in the following we take $f_k$  as implicitly defined in Eq. (\ref{delta}) and the PTRG flow equation has a particularly simple form
\be\label{fulleq}
k\partial_k S_k={\rm Tr}~{\rm exp} \left \lbrack-\frac{S_{k,(\sigma\sigma)}^{''}-2 S_{k,(\sigma\psi)}^{''}~ {\cal A}_k^{-1} ~S_{k,(\overline\psi \sigma)}^{''} }{Z_k k^2}
\right \rbrack -{\rm Tr}~{\rm exp} \left \lbrack -\frac {{\cal A}_k^\dagger {\cal A}_k }{(Z^{\psi}_k k)^2}\right \rbrack
\ee
where we used a concise notation for the second derivatives of the action and $Z_k$ and  $Z^{\psi}_k$ provide the correct normalization of the two exponential terms. 
The traces are extended to all the relevant degrees of freedom.

The extension of the flow equation to finite temperature is done following the standard procedure \cite{kapu}, i.e. by imposing periodic boundary conditions on  the Euclidean 
time component with period fixed by the inverse temperature $1/T$ and consequently discretizing the corresponding component of the momentum.  
Therefore the integration on the component $q_4$ (the index $4$ is taken in analogy to the one used for the gamma matrix $\gamma^4$) is to be 
converted into the infinite sum over the Matsubara frequencies, which are respectively $q_4 \to \omega_n= 2 n\pi  T$ for bosons and  
$q_4 \to \nu_n=( 2n+1) \pi T$ for fermions and  $n \in \Z$, $\int d q_4 \to 2\pi T \sum_n $. 

Finally, finite fermion density effects are accounted for by adding to the bare action the chemical potential $\mu$ coupled to the density $\psi^\dagger \psi$.
Then we can compute the purely fermionic contribution to Eq. (\ref{fulleq}) in the momentum space 
\be\label{acrocea}
{\cal A}_k^\dagger {\cal A}_k=(Z^{\psi}_k)^2 (q^2 +  \nu_n^2) +\mu^2 +g_k^2 \sigma^2 -2i\mu ( Z^{\psi}_k q \hskip -0.2 cm \slash + g_k \sigma) \gamma^4 
\ee
where $N_f$ has been absorbed into the coupling constant $g_k \sqrt{N_f}=\widehat g_k$. 

We are now able to extract from the full flow equation (\ref{fulleq}) the set of coupled equations for the various $k$ dependent terms of  the effective action.
The equation for the potential $U_k$ is simply obtained by considering constant field configurations for $\sigma$ and vanishing fermion fields in Eq. (\ref{fulleq})
and, by performing the integral on  the momentum,  we get 
\be\label{pote}
k\partial_k U_k=\Omega_b + \Omega_f
\ee
where the boson contribution, after performing the momentum integrals, is
\be\label{boso}
\Omega_b= \frac{T k^2}{4\pi} \sum_n {\rm exp}\left \lbrack -\frac {\omega_n^2}{k^2} - \frac{ M_b } {Z_k k^2} \right \rbrack
\ee
with  $M_b=2\dot U_k+4\sigma^2   \ddot U_k$ (the dots indicate derivatives with respect to $\sigma^2$), 
and the fermion contribution is 
\bea
&&\Omega_f= -2 N_f T \sum_n \int \frac{d^2 q}{(2\pi)^2} ~
{\rm exp}\left \lbrack -\frac {\nu_n^2}{k^2}  \right \rbrack \times \nonumber\\
&&
\left \lbrace  {\rm exp}\left \lbrack -\frac{ \left ( \mu - \sqrt{{Z^{\psi}_k}^2 q^2 +g_k^2\sigma^2}\right )^2 } {(Z^{\psi}_k k)^2 } \right \rbrack + 
 {\rm exp}\left \lbrack -\frac{ \left ( \mu + \sqrt{{Z^{\psi}_k}^2 q^2 +g_k^2\sigma^2}\right )^2 }{(Z^{\psi}_k k)^2}   \right \rbrack  \right \rbrace
\label{fermi}\eea
Equation  (\ref{fermi}) shows the fermion contribution in a compact form. However the momentum integrals can be performed analytically which 
makes easier the numerical integration of the set of flow equations.

In  order to derive the equation for $Z_k$ one has to expand the exponentials in the right hand side of Eq. (\ref{fulleq}) retaining the commutators of the coordinate and 
momentum dependent terms, and collecting the full coefficient of $\partial_\mu  \sigma \partial_\mu  \sigma$. 
This procedure is already discussed in \cite{oldza1} and, specifically for  the
PTRG flow equation of $Z_k$ in scalar theories, in \cite{boza,maza,zappa1}. Moreover here we are only interested in   the leading  corrections to the flow of the
potential $U_k$ and therefore,  instead of solving the problem by taking  $Z_k$,  as well as $Z^\psi_k$ and $g_k$, as arbitrary functions of the fields, we 
we limit ourselves to consider the particular value of these parameters  when computed at the minimum of the potential at each value of $k$, $\sigma=\overline \sigma_k$
and for vanishing fermion fields. The lowest order equation  for $Z_k$ is therefore:
\bea
&& k\partial_k Z_k =
 -\frac{T k^2}{2\pi} \sum_n 
{\rm exp}\left \lbrack -\frac {\omega_n^2}{k^2} - \frac{ M_b } {Z_k k^2} \right \rbrack
~~\frac{2\overline \sigma_k^2 Z_k 
(\dot M_b)^2}{3 (Z_k k^2)^3} \nonumber\\&&
-\frac{T N_f g_k^2 }{(Z_k^\psi k)^2 \pi }  \sum_n  
{\rm exp}\left \lbrack -\frac {\nu_n^2}{k^2} - \frac{ \mu^2+ g_k^2\overline\sigma_k^2 } { ({Z^{\psi}_k} k)^2 } \right \rbrack
\left ( 1-\frac{4 g_k^2\overline\sigma_k^2}{3 (Z^{\psi}_k k)^2} \right )
\label{zetsca}
\eea

The evolution equation  of $Z^\psi_k$ is obtained by taking   the coefficient of $\overline \psi_j \partial \hskip -0.2 cm \slash \psi_j$
 after expanding the exponentials in the right hand side of Eq. (\ref{fulleq}):
\bea
&&k\partial_k Z^\psi_k= \frac{2 g_k^2}{Z_k k^2}  T \sum_n \int \frac{d^2 q}{(2\pi)^2}~ 
{\rm exp}\left \lbrack -\frac {  Z_k (q^2+\omega^2_n) + M_b }{Z_k k^2} \right \rbrack \times
\nonumber\\
&&\left( 
\frac{2 {Z^\psi_k}^2 (q^2+\nu_n^2 )}{3} 
~\frac{W^2- 4 {Z^\psi_k}^2\mu^2\nu_n^2}
{\left( W^2+ 4 {Z^\psi_k}^2\mu^2\nu_n^2\right)^2  }
 -\frac{W}{W^2+ 4 {Z^\psi_k}^2\mu^2\nu_n^2} 
\right )
\label{zetfer}
\eea
and the one for $g_k$ by selecting the coefficient of $\sigma \overline \psi_j \psi_j$:
\be\label{gi}
k\partial_k g_k= \frac{2 g_k^3}{Z_k k^2}  T \sum_n \int \frac{d^2 q}{(2\pi)^2}~  
{\rm exp}\left \lbrack -\frac { Z_k (q^2+\omega_n^2) + M_b  }{Z_k k^2} \right \rbrack 
~\frac{W}{W^2+ 4 {Z^\psi_k}^2\mu^2\nu_n^2}
\ee
where $M_b$ and $\dot M_b$ in Eqs. (\ref{zetsca}),(\ref{zetfer}),(\ref{gi}) are evaluated at $\sigma=\overline\sigma_k$
and  $W$ is defined as 
\be\label{ausi}
W={Z^\psi_k}^2  (q^2+\nu_n^2 ) +g_k^2\overline\sigma_k^2- \mu^2
\ee

However, in the following we shall not solve the full  potential equation Eq. (\ref{pote}) which is a partial differential equation, but rather we  treat a set of ordinary differential 
equations for the various coefficients of a  truncated  polynomial expansion of the potential.  In fact, according to \cite{pirner2} such truncations converge quite rapidly and 
we choose to retain powers of the field up to $\sigma^8$ in the expansion. Therefore we parametrize the potential, expanded around the local minimum $\overline \sigma_k$,  
in the following way
\be\label{expa}
U_k(\sigma^2)= u_{0,k}+\frac{1}{2} u_{2,k} (\sigma^2- {\overline \sigma_k}^2) +\sum_{i=2}^4 u_{2i,k} (\sigma^2- {\overline \sigma_k}^2)^i 
\ee
The parametrization in Eq. (\ref{expa}) contains five $k$ dependent parameters. In fact  when the minimum of the potential is located at $\overline \sigma_k=0 $, then 
the independent parameters are $u_{0,k},~u_{2,k},~u_{4,k},~u_{6,k},~u_{8,k}$. When instead the minimum is at a nonvanishing $\overline \sigma_k$,  one has 
$ {\dot U_k}(\overline \sigma^2)=0$ and then  $u_{2,k}=0$ identically and the  $k$ dependent parameters are still five with $u_{2,k}$ replaced by $\overline \sigma_k$. 
For our potential at $k=\Lambda$ we have   $u_{2,\Lambda}=1$ and the minimum at $\overline \sigma_\Lambda=0$ and therefore we start our flow
by explicitly considering the evolution of $u_{2,k}$.
Then,  when $k$ is lowered,  $u_{2,k}$  diminishes and (only for suitable values of $T$ and $\mu$) eventually vanishes  at $\overline k$
( with $0< \overline k<\Lambda$). So, at $k=\overline k$, we replace the flow equation for $u_{2,k}$ with the one for $\overline \sigma_{k}$, with  the initial condition  
$\overline \sigma_{\overline k}=0$ and require continuity for the other parameters  $u_{0,k},~u_{4,k},~u_{6,k},~u_{8,k}$.

The flow equations for the parameters in Eq. (\ref{expa}) are directly obtained from   Eq. (\ref{pote}) by repeated differentiations with respect to $\sigma^2$
(we shall not display them explicitly) and the one for $\overline \sigma_k$  is instead derived from the minimum condition,
\be\label{eqsig}
(2 \overline \sigma_k \ddot U_k (\overline \sigma_k^2))k \partial_k \overline \sigma_k= -k \partial_k \dot U_k (\overline \sigma_k^2)
\ee
Incidentally we note that $u_{0,k}$ never appears in the other equations and therefore this parameter does not affect the evolution of the others.

As a first step in the analysis of the flow equations, we briefly review the well known case  $N_f\to \infty$,  that can be treated analytically. In fact, since  the minimum of the 
potential increases as $\sqrt { N_f}$ while  the generated mass of the fermion does not depend on $N_f$ (we recall that in our definitions the coupling $g_k$ is proportional to 
$1/\sqrt { N_f}$ ) it is convenient to rescale the fields as $\sigma \to \sqrt { N_f} \sigma$,  $\psi\to \sqrt { N_f} \psi$ as well as the potential $U_k \to N_f U_k$ so that the action,
expressed in terms of these new variables and of $\widehat g_k$,  is homogeneously rescaled by the  factor $N_f$. 
It is easy to realize that after these changes the right hand sides of the flow equations for $Z_k^\psi$ and $\widehat g_k$ are suppressed in the limit  $N_f\to \infty$ 
and these parameters remain fixed to their initial value at $k=\Lambda$ and in particular $Z_k^\psi=1$.  
Also, in Eq. (\ref{pote}) the term $\Omega_b$  is suppressed and the equation for the potential depends only on the fermionic contribution which does not contain any 
dependence on the potential itself or on its derivatives
or on the parameter $Z_k$.  
As a consequence, the equations for $Z_k$ and for $U_k$ decouple. We neglect the former and focus on the latter, which is relevant for the symmetry breaking problem,
and we note that  in this limit  Eq. (\ref{pote}) is no longer a differential equation and can be simply integrated in $k$ from zero to infinity with the help of Eq. (\ref{loga}) and 
by recalling the boundary condition for the potential in the ultraviolet region assigned in Eq. (\ref{actionone}). 
By extending the $k$ integration up to infinity,  we have implicitly  removed the ultraviolet scale $\Lambda$ and this will introduce some divergences in our computation which
at some point  need to  be regulated. However in this way we can more easily make contact with the standard calculations of the large $N_f$ case (see e.g.  \cite{cooper}). 
In order to determine the extrema of the potential it is convenient to focus on its derivative with respect to  $\sigma$ and, after performing the sum over the Matsubara 
frequencies \cite{kapu}, we get ( we neglect for simplicity the subscript $k$ for the coupling $\widehat g$ since here, as noticed above, it is $k$ independent ) 
\be\label{derpot}
\frac{\partial U_{k=0}}{\partial \sigma} = \sigma - \int \frac{d^2 q}{(2\pi)^2}~\frac{\widehat g^2 \sigma}{\sqrt{q^2+\widehat g^2 \sigma^2} } \left \lbrack 
{\rm tanh} \frac{\sqrt{q^2+\widehat g^2 \sigma^2}+ \mu}{2 T} + {\rm tanh} \frac{\sqrt{q^2+\widehat g^2 \sigma^2}-\mu }{2 T} \right \rbrack  
\ee

In the particular case $\mu=T=0$,  
if we  regulate in Eq. (\ref{derpot}) the divergent  integral on the momentum $q$ with the ultraviolet  cutoff $\Lambda$, we get the gap equation 
that defines the fermion mass at zero temperature and chemical potential \cite{cooper}
\be\label{gap}
\frac{\pi}{\widehat g^2}=\Lambda -\widehat g \sigma=\Lambda -m_f
\ee

If we consider $T=0$ and finite chemical potential we see that the hyperbolic tangents in Eq. (\ref{derpot}) in the limit of vanishing temperature become step functions and,
by making use of the gap equation (\ref{gap}) to get rid of the regulator $\Lambda$,  it is easy to check that taking  $\mu<m_f$ or  $\mu>m_f$  yields two different behaviors. 
In fact in the former case the same gap equation as for $\mu=0$ is obtained, in the latter case we get the  extremum condition $\widehat g^2 \sigma (\mu -m_f) =0$ that 
has the unique solution $\sigma=0$. Therefore at $T=0$ a first order transition is observed when the chemical potential is equal to $m_f$. 

Finally when also $T\neq 0$, again with the help of Eq. (\ref{gap}), we get the gap equation
\be\label{muet}
\left ( {\rm cosh}\frac{\widehat g \sigma}{T}- \frac{ e^{(m_f/T)}}{2} +{\rm cosh}\frac{\mu}{T} \right )\sigma=0
\ee

For nonvanishing $T$, Eq. (\ref{muet}), in addition to $\sigma=0$,  has another solution only for sufficiently small chemical potential $\mu<\overline \mu(T)$ ( with 
$\overline \mu(T) \sim m_f$ for small $T$ ). This solution tends to zero for $\mu \to \overline \mu(T)$ from below 
and disappears for $\mu>\overline \mu(T)$ where the only extremum is 
$\sigma=0$. Then the transition in this case is second order. Only at $T=0$ there is a  singular behavior that leads to a first order transition \cite{hands2}.

Let us now consider the finite $N_f$ case and, in order to test the PTRG we shall compare our analysis with the lattice results obtained in \cite{hands2} where $N_f=12$ is 
taken to observe the leading corrections to the $N_f\to\infty$ case. 
Therefore in the following we select  this particular value of $N_f$ and solve numerically our  flow equations. The flow starts  at the scale $k=\Lambda$ and ends at $k=0$ 
and we varied $\Lambda$ in the range $10^3 -  10^5$ to test the insensitivity of the final output of the dimensionless parameters to the specific value of the ultraviolet scale.
The initial values of the parameters at $k=\Lambda$ are those indicated below Eq. (\ref{effact}) and it must be noted that $u_{2,\Lambda}=1$, independently of the particular 
value of $\Lambda$ and this implies that  $u_{2,k}$ must be dimensionless. Consequently all the canonical dimensions of the other parameters must be arranged according 
to this  point.
 
The most simple version of the flow equations is obtained by a further reduction of the number of  running parameters and, specifically, by neglecting the flow of the 
renormalization $Z_k$ of the scalar field. Since in our problem the bare value of this parameter is $Z_\Lambda=0$, this approximation corresponds to have  $Z_k=0$
at  any scale $k$ and therefore to  totally neglect the kinetic contribution of the scalar field. Moreover, as it can be seen from Eqs. (\ref{zetfer}) and (\ref{gi}), in this case 
also $Z^\psi_k$ and $g_k$ become $k$-independent and stay fixed to their initial value and in Eq. (\ref{pote}), $\Omega_b=0$. When $Z_k$ is turned on and the full set 
of flow equations is considered, still $Z_k^\psi$ and $g_k$ do not show any significant change from their initial values.
On the contrary $Z_k$ has a not negligible  flow and when the scale $k$ goes from the ultraviolet to the infrared region, this parameter grows from zero to a finite value, 
and this  was already observed in models of fermions coupled to scalars \cite{wett1,oldza2}. 
In general  the effects of the scalar renormalization are significant, as it happens for instance in the quantitative determination of the critical exponents of the scalar theory
or of some  energy levels in quantum mechanics \cite{maza,zappa2} and, also in this problem they will turn out to be relevant.

At $\mu=T=0$ and with $N_f=12$ (which is the value of $N_f$ adopted in  \cite{hands2})
we find a critical value of the coupling  that separates the symmetric and  the broken phase, which, in cutoff units, is 
$1/ ({\widehat g_\Lambda^*} \sqrt{\Lambda})^2=0.178(2)$. In  \cite{hands2} the inverse squared critical coupling, expressed in different units, is $0.98$ and, by addressing 
the difference between these two determinations to the two different scales employed  to normalize the dimensionful coupling, it follows that the ratio of these two numbers 
directly yields  the ratio of the two scales. Then, since  the value of the inverse square coupling taken in  \cite{hands2}
to study the restoration of the chiral symmetry at finite $T$ and $\mu$ is $0.75$, we can  easily convert it  in units of our cutoff $\Lambda$
to get the input value of the coupling in our flow equations. In our scale it is $1/ (\widehat g_\Lambda \sqrt{\Lambda})^2=0.136$. 
At this value of the coupling and again at $\mu=T=0$ we have determined the fermion mass which, according to the criterion chosen  in  \cite{hands2}, will be the parameter
used to normalize the dimensionful quantities in our analysis at finite $\mu$ and $T$. 
In particular we found $m_f/\Lambda=0.146$ and, when  $Z_k=0$, $m_f/\Lambda=0.145$.

The phase diagram found by means of the flow equations is shown in Fig. 1.
The solid line shows the transition at $N_f\to\infty$ as obtained from Eq. (\ref{muet}).
Except for the points at $\mu=0$ where the temperature is raised in order to observe the symmetry restoration, all the other points are obtained by keeping a fixed 
temperature and changing the chemical potential $\mu$ to induce the transition and  the  values of the temperature considered are: 
$T/m_f=0,~0.24,~0.29,~0.36,~0.48$.  For the circles at $\mu=0$ 
and at $T/m_f=0.48$ the errors are small and they are not displayed  in the figure. Notice that the circle at $\mu=0$ 
is almost coincident with the lattice determination of \cite{hands2} which corresponds to the star (together with its  error bar) on the $T$ axis in Fig. 1. 
As expected the circles are in very good agreement with the mean field line and it is necessary to go beyond the local potential approximation 
to improve the results. The inclusion of the flow of  $Z_k$ induces  a significant change, reducing the  values of $T$ and $\mu$ at which the transition occurs.
This effect, qualitatively correct, is somehow large and the points turn out to be systematically lower than the lattice results.

Also,  some uncertainties affect our determinations of the transition points in the $\mu-T$ plane  as it is shown by the error bars 
associated to the  points in Fig. 1.  In fact,  for values of $\mu$ below these  bars ( or, in the case of  the transition  at $\mu=0$, 
for values of $T$ below the  error bar) we find the absolute minimum of the potential at a nonvanishing vacuum expectation of the field whereas,
for values above these  bars the minimum corresponds to a vanishing vacuum expectation of the field and  chiral symmetry is thus restored.
For values of $\mu$ (and $T$) within the  bars  it is difficult to determine the nature of the transition, especially if  the flow 
of $Z_k$ is included and therefore, by taking the most  conservative point of view, we explicitly indicated in Fig.1 the ranges of $\mu$ and $T$
where these uncertainties are present. 

As an example, in Fig. 2 $\overline\sigma_{k=0}/m_f$ is displayed as a function of $T/m_f$. 
The solid line is obtained for  $Z_k=0$  and it shows a continuous transition and the dashed line corresponds to $Z_k\neq 0$. 
The latter curve has a discontinuous fall to  $\overline\sigma_k=0$ which should indicate a first order transition.
But in the temperature range indicated by  the error bar on the $x$-axis of Fig. 2  (which is the same error bar shown  for  the corresponding point in Fig. 1) 
we observe in our flow equations the appearance,  at small finite $k$, of a non-vanishing minimum of the potential which however, still at finite $k$, vanishes.
Then, for temperatures above the error bar in Fig. 2 the minimum of the potential corresponds to $\overline\sigma_k=0$ at any value of $k$.
Because of this effect that , incidentally,  has already been observed in  \cite{pirner2}, we believe that 
it is not safe to conclude that the inclusion of the flow of $Z_k$ has  changed the order of the transition which moreover would be in contrast with the expectation of a 
second order transition. Rather, we would argue that it is more likely that  a different truncation in our flow equations, which  includes more operators, like field dependent 
contributions to the scalar wave function renormalization, could restore the smooth behavior obtained when $Z_k=0$ and the dashed line in Fig. 2 would continuously 
reach zero as it happens for the other one.

In conclusion the PTRG flow equations provide a good qualitative picture of the transition at finite temperature and chemical potential, and in particular they have the 
correct behavior in the limit $N_f\to\infty$. Also Fig. 1 shows that the  role of $Z_k$ is certainly relevant in the determination of the critical line and a reasonable 
comparison with the phase diagram obtained from the  lattice simulations in \cite{hands2}, is possible. As a further check we have computed, following \cite{hands2},  
the exponent $\nu$ for the critical  behavior around  the coupling $g^*$ at  $T=0$, and we found $\nu=1.06(4)$  when $Z_k=0$ and $\nu=1.19(30)$  for the full set of 
equations, which has to be compared to the value $\nu=1.05(10)$ found in  \cite{hands2}. Again the inclusion of $Z_k$ gives a correction larger that the one expected.
On the other hand, the  analysis of the order of the phase transition at this level of approximation is not satisfactory, although some expected features are recovered.
In our opinion the quantitative discrepancies as well as the various uncertainties discussed have to be addressed to the specific truncations and approximations made 
on the flow equations. We expect that  the inclusion of field dependent terms in  the scalar wave function renormalization (as it has been already pointed out  in 
\cite{deandrea} in a different context ) could significantly improve the analysis and  therefore this turns out to be an essential ingredient  to carry out  an accurate analysis of 
phenomenologically realistic models.

\eject

\begin{figure}
\epsfig{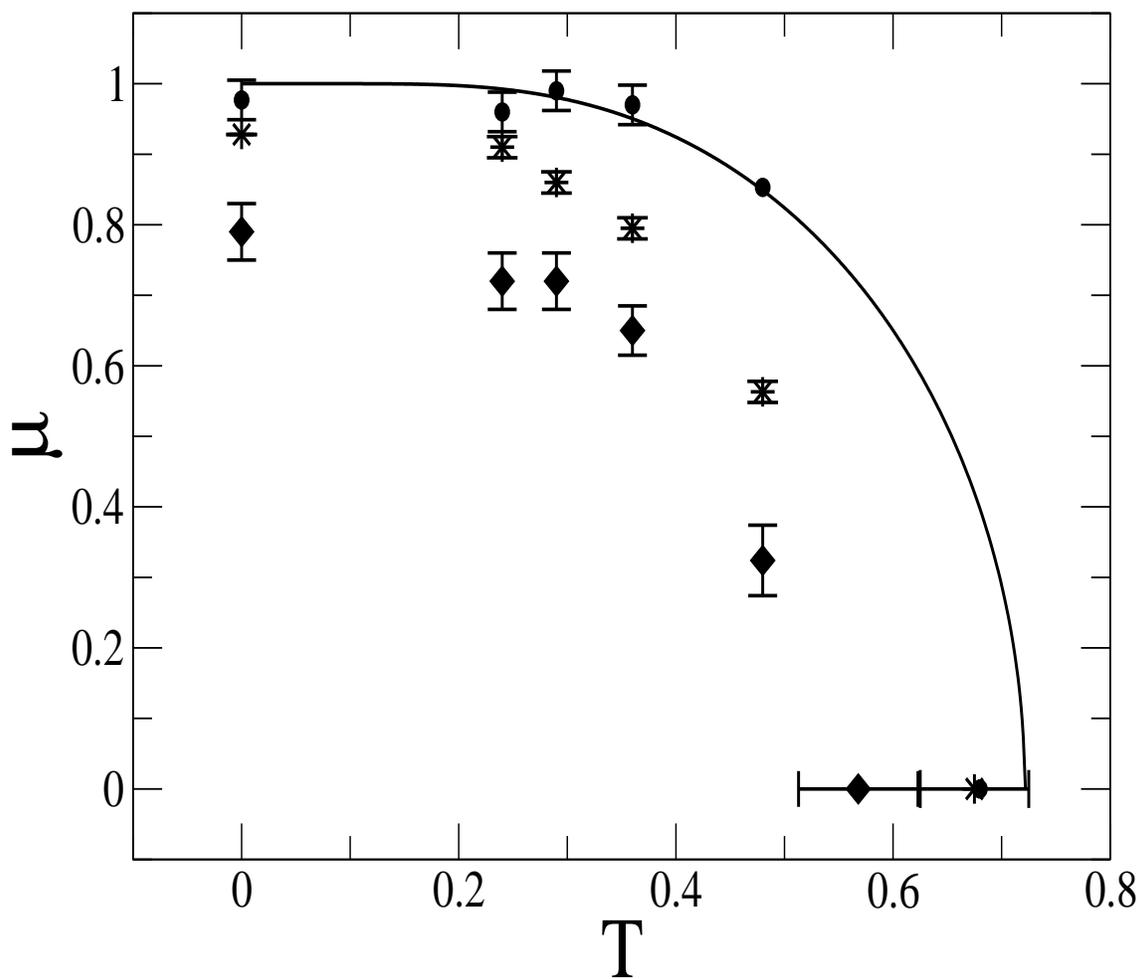}
\caption{ 
Critical line on the plane $T-\mu$. The solid line corresponds to the transition for 
$N_f\to\infty$. The circles are obtained from the flow equations with $Z_k=0$ and the 
diamonds for $Z_k\neq 0$. The stars correspond the lattice results as quoted in \cite{hands2}. 
$T$ and $\mu$ are expressed in units  of the fermion mass $m_f$.
}
\end{figure}

\eject

\begin{figure}
\epsfig{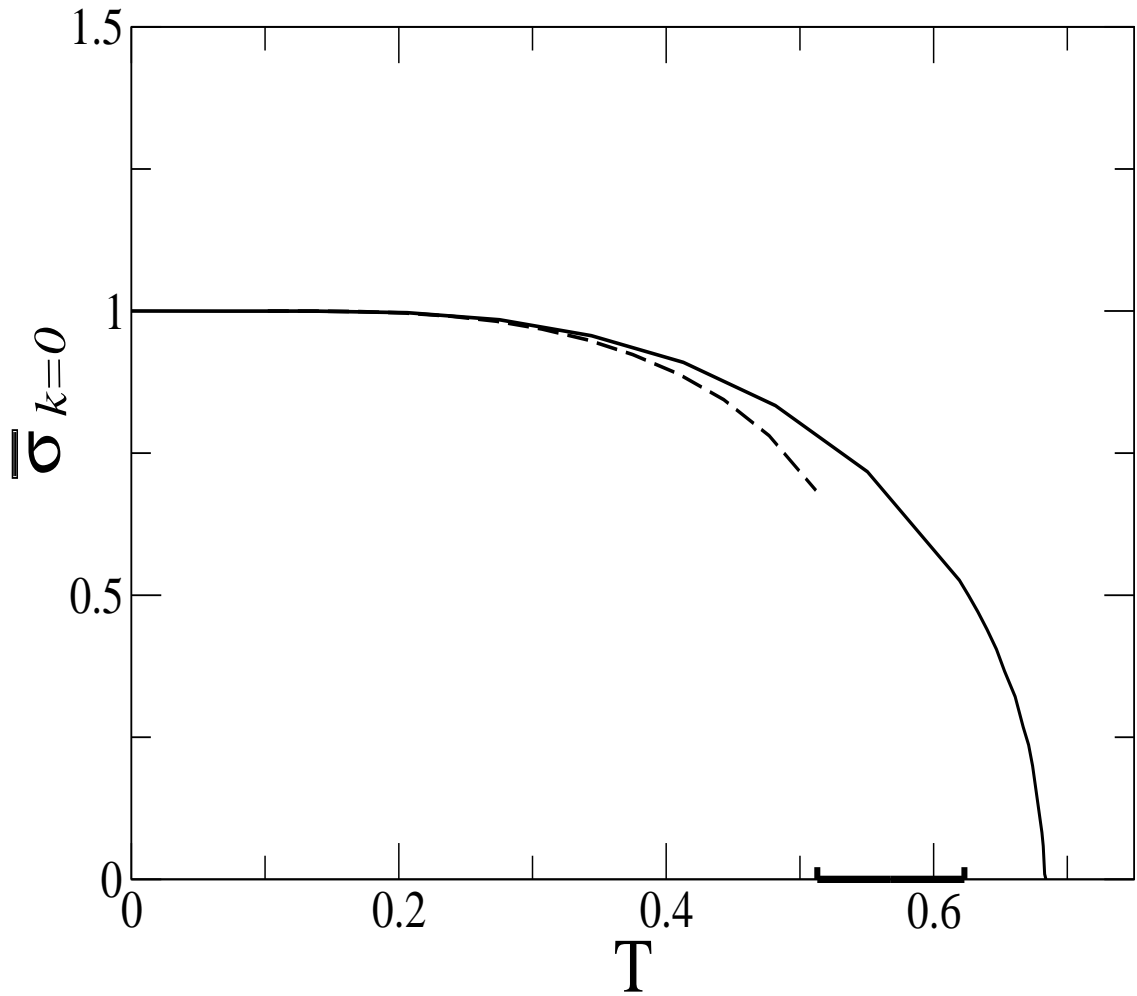}
\caption{
$\overline\sigma_{k=0}$ vs. $T$ ( both quantities in units of $m_f$). 
The solid line corresponds to $Z_k=0$ and the dashed one to $Z_k\neq 0$. 
For the error bar on the $x$-axis see text.
}
\end{figure}

\end{document}